# A NEW TASK FOR PHILOSOPHY OF SCIENCE

NICHOLAS MAXWELL

**Abstract:** This paper argues that philosophers of science have before them an important new task that they urgently need to take up. It is to convince the scientific community to adopt and implement a new philosophy of science that does better justice to the deeply problematic basic intellectual aims of science than that which we have at present. Problematic aims evolve with evolving knowledge, that part of philosophy of science concerned with aims and methods thus becoming an integral part of science itself. The outcome of putting this new philosophy into scientific practice would be a new kind of science, both more intellectually rigorous and one that does better justice to the best interests of humanity.

## 1. Introduction

We philosophers of science have before us an important new task that we urgently need to take up. It is to convince the scientific community to adopt and implement a new philosophy of science that does better justice to the deeply problematic basic intellectual aims of science than that which we have at present. The outcome of putting this new philosophy into scientific practice would be a new kind of science, both more intellectually rigorous and one that does better justice to the best interests of humanity. It will be, I admit, a hard task to convince the scientific community that the conception of science they tend to take for granted needs to be radically improved. Not all scientists hold the philosophy of science in high esteem. We need, nevertheless, to do what we can to get across the argument that a new conception of science needs to be adopted and implemented, one that acknowledges and seeks to improve problematic aims of science.

## 2. The Problematic Aims of Science

The crucial question that has to be decided to determine the extent to which the philosophy of science has fruitful implications for science itself is simply this: Are the basic aims of science unproblematic and fixed? Or are they problematic, permanently in need of improvement, so that the aims and associated methods of science need to *be improved* as scientific knowledge itself improves? If the former holds, there is no reason why the activity of articulating the aims and methods of science—the philosophy of science, in other words—should have much of an impact on science itself.[1] The task is simply to get clear what the basic aim of science is, and what the basic methods are, once and for all. This task may well be what might be called a "meta" task, one that leaves science itself unaffected. The task is to improve our knowledge and understanding of the nature of the scientific enterprise, but not to change science itself. That the philosophy of science, conceived in these terms, has nothing much to offer science itself is no cause for shame.

---

[1] Philosophy of science does much more than seek to articulate the aims and methods of science, and justify them. In what follows I am primarily concerned with that part of philosophy of science concerned with the aims and methods of science.

But the situation is radically changed if the aims of science are inherently problematic, permanently in need of improvement as scientific knowledge and understanding improve, so that there is something like positive feedback between improving scientific knowledge and improving aims and associated methods—improving knowledge about how to improve knowledge. Philosophy of science—construed as the enterprise of articulating and justifying the aims and methods of science—becomes a vital, integral part of science itself. As science evolves, it influences the philosophy of science; and vice versa, as philosophy of science evolves, it influences science. Or at least that is what ought to go on.

Let us give a bit more flesh to these very abstract considerations. I now formulate two rival philosophies of science that give starkly opposing answers to the above question. The first holds that science has a fixed aim, and fixed associated methods; the second that science has profoundly problematic, and so evolving, aims, and thus evolving associated methods. The first implies that philosophy of science is a meta-discipline, one that leaves science essentially unaffected—just as astronomical study of the moon leaves the moon unaffected. The second implies that science and philosophy of science interact with each other in both directions.

But first, a few words about the aims of science. Our concern here is very definitely with the aims of *science*, not the aims of scientists. Furthermore, our concern is not with what the scientific community *declares* to be the aims of science; rather, what matters is aims that are implicit in the actions of the scientific community—aims that, it seems, scientific actions strive to attain. Questions about the aims of science arise in two contexts: the context of discovery and the context of justification—the context of determining what is, and what is not, a contribution to science. Our concern here is primarily with the aims of science in the latter context.

How are rival proposals as to what constitutes an aim for science to be assessed? What kinds of consideration can be brought to bear in deciding whether such and such a proposed aim is acceptable? I suggest that there are at least the following three considerations.

1. The proposed aim must be such that it does reasonable justice to relevant actual scientific actions—acceptance and rejection of potential contributions to science. It must be possible, in other words, to interpret these actions, in a reasonably good way, as attempts to realize the aim in question—so that attributing the aim to science makes reasonably good sense of what scientists do in practice. (We do not require that the aim make sense of everything scientists do, in every detail and respect.)
2. The proposed aim must be sufficiently worthwhile. It must do justice to the intellectual value of science.
3. It must be possible to show that scientific actions—contributions that are accepted—do indeed constitute steps towards realizing the aim, or at least that there are sufficiently good reasons to hold that this is the case. The aim is, in other words, realizable, as far as we know, and contributions to science do indeed constitute steps towards realizing the aim, or at least we have reasonably good reasons to hold that this is the case.

The third of these three requirements is the most controversial. In order to meet it, we need to solve the problem of induction. I will have more to say about this issue below.

Here, now, are my two rival philosophies of science. They are, more accurately, two rival philosophies of physics; or perhaps even more accurately, two rival philosophies of theoretical physics.

*Standard empiricism.* The basic aim of physics is truth, nothing being presupposed about the truth. The basic method is to assess claims to knowledge impartially with respect to evidence. When it comes to deciding what laws and theories to accept, considerations of simplicity, unity, or explanatory power may legitimately influence choice of law or theory in addition to empirical success and failure, but not in such a way that nature herself, or the phenomena, are presupposed to be simple, unified, or comprehensible. Again, choice of theory may, for a time, be biased in the direction of some metaphysical view, Kuhnian paradigm, or Lakatosian "hard core." The decisive point is that *no permanent thesis about the world can be accepted as a part of scientific knowledge independent of evidence*.[2]

*Aim-oriented empiricism.* The basic aim of physics is to discover the truth *presupposed to be physically explanatory* or *comprehensible*. That is, the basic aim of physics presupposes that an underlying, unified, physical something exists in nature, inherent in all phenomena, that suffices to explain everything that occurs. This aim is profoundly problematic because it is a sheer metaphysical conjecture that the universe is comprehensible in the sense indicated. The basic method comes in two parts. Part 1 subjects the profoundly problematic metaphysical conjecture about the physical comprehensibility of the universe, inherent in the basic aim of physics, to sustained critical scrutiny, rival possible versions of the conjecture being articulated and assessed, that version of the conjecture being provisionally accepted that leads, or promises to lead, to the most empirically progressive research programme. Part 2 involves accepting those theories that are sufficiently in accord with (a) evidence and (b) the best available metaphysical conjecture concerning the physical comprehensibility of the universe.[3]

Let us put to one side, for the moment, objections to these two views, questions about which is to be preferred, or disliked least. The key point I want to emphasize is that these two rival philosophies of physics have very different implications for how physics and the philosophy of physics ought to be inter-related.

Given standard empiricism, with its fixed aim for physics, and with its broadly fixed methods, there is not much room for that fragment of philosophy of physics concerned with the aims and methods of physics to have much of an impact on physics itself. The task of the philosophy of physics is to make explicit what is presumably implicit in scientific practice: make explicit the basic aim and the methods adopted in pursuit of that aim, and then provide a justification for the methods in question. The latter involves at least demonstrating that the methods in question are the best to adopt granted that one seeks to realize the specified aim. None of this has implications for how physics itself should be conducted—not unless physicists fail in practice to put the specified methods into practice. Physics and the philosophy of physics, granted standard empiricism, seem sharply distinct. This is indeed declared to be the case by standard empiricism itself. The basic method of standard empiricism is to assess claims to knowledge by means of evidence. Anything that is not open to being so assessed cannot be a part of science. But the philosophy of science—or, more specifically, the philosophy of physics—is a view about what the basic aim and methods of physics ought to be. In being concerned with norms or ideals, it is not the sort of entity that is empirically testable. Thus, granted standard empiricism, the philosophy of physics is not a part of physics itself. This conclusion is borne out, in particular, by Karl Popper's demarcation requirement, which holds that a discipline, in order to be scientific, must consist of empirically falsifiable theories or statements (Popper 1959, 40–42). The philosophy of physics does not consist of such theories; hence it is not itself a part of physics. Standard empiricism itself drives a wedge between science and the philosophy of science.

---

[2] This rather meagre thesis of standard empiricism is common ground for such otherwise diverse doctrines as logical positivism, inductivism, logical empiricism, hypothetico-deductivism, conventionalism, constructive empiricism, pragmatism, realism, inference-to-the-best-explanationism, the views of Popper, Kuhn, and Lakatos, and many more recent views as well: see Maxwell 1998, 38–45; 2004, chap. 1, nn. 5, 6, and 14). For discussion of the claim that Popper, Kuhn, and Lakatos defend versions of standard empiricism, see Maxwell 2005.

[3] For expositions of aim-oriented empiricism, see Maxwell 1974; 1998; 2004; 2005; 2007, chap. 14; 2013; 2017a; 2017b; 2017c; 2019a, chap. 3).

All this changes dramatically if we move from standard to aim-oriented empiricism. According to aim-oriented empiricism, there are profoundly problematic metaphysical theses about the nature of the universe inherent in the aims of physics. These presupposed theses, in the more or less specific form that they are accepted at any stage in the development of physics, are almost bound to be false. And yet—according to aim-oriented empiricism—they exercise a profound influence over the kind of theory that is sought in the context of discovery, and the kind of theory that is accepted in the context of justification. If physics is to meet with success, it needs to make good choices of basic metaphysical assumptions, and yet it is here that physics is almost bound to get things very seriously wrong, since we are concerned with mere unfounded *conjectures*—wild *speculations*—about that of which we are most ignorant, the ultimate nature of the universe. And if we glance at the past, we see that physics has indeed got things very wrong, again and again, in that it has blundered from one grossly false metaphysical speculation to another. In the seventeenth century there was the idea that the universe is made up of corpuscles that interact by contact. This gave way to the Boscovichean idea that it is made up of point particles with mass, surrounded by rigid spherically symmetrical forces that act at a distance; this in turn gave way to the idea that it is made up of charged point particles embedded in a classical field, which in turn became the idea that there is just one self-interacting, unified field. This became the idea that the universe is compounded of quantum entities (whatever they may be), which in turn became a unified quantum field, and then quantum strings in ten or eleven dimensions of space-time. To repeat: if physics is to meet with success, it must make a good choice of basic metaphysical assumption, but it is just here that it is bound to get things drastically wrong. The only hope—since we cannot avoid making *some* assumption—is to subject such assumptions to sustained critical scrutiny and development, in an attempt to improve the specific assumption that is adopted.

This crucial activity of imaginatively developing and critically assessing metaphysical assumptions for physics in an attempt to improve the specific assumption that is adopted is thus likely to have a profound influence over theoretical physics itself, in the contexts of both discovery and justification, whether for good or for ill. If physics makes a good choice of metaphysical conjecture, it may well forge ahead; if it makes a bad choice, it will be stultified. But this enterprise of exploring possible metaphysical assumptions for physics is *the philosophy of physics*. It is concerned with empirically untestable, metaphysical ideas. It seeks to improve metaphysical conjectures implicit in the basic aim of physics, thus improving the permanently problematic aim. And in seeking to improve metaphysical assumptions, it seeks also to improve associated *methods*.

Thus, according to aim-oriented empiricism, the philosophy of physics—or that central part of the philosophy of physics concerned with the aims and methods of physics—is (or ought to be) a vital, integral part of physics itself, exercising a very substantial influence over physics in the contexts of both discovery and justification.[4]

## 3. A Fundamental Watershed for Science and Philosophy of Science

The key difference between standard empiricism and aim-oriented empiricism has to do with whether the basic aims of science do not, or do have problematic assumptions inherent in them. Standard empiricism declares that the basic intellectual aim is factual truth. How we get it, how we can know whether or not we have got it: these questions are profoundly problematic. But the aim itself is not problematic. It is fixed: factual truth in so far as it can be attained.[5] "Snow is white" is true if and only if snow is white: that is all there is to it. There are no substantial assumptions inherent in the aim all too likely to be false that need persistent revision and improvement.

Aim-oriented empiricism, by contrast, holds that there is a substantial and profoundly problematic metaphysical assumption built into the basic aim of physics: the universe is physically comprehensible. It is such that there is a true physical "theory of everything" that is *unified*. This assumption, and more specific versions of this assumption, may well be false. As we have seen, the big, problematic assumption inherent in the aim of physics may well need to be modified and improved, again and again, as physics proceeds.

This key dividing line between standard and aim-oriented empiricism constitutes a quite fundamental *watershed* for science, for philosophy of science, and for the relationship between the two. Stand anywhere on the standard empiricist side of the watershed (upholding any one of the well-known views about science) and one will hold that science has a fixed basic intellectual aim and fixed methods.[6] Science and philosophy of science are distinct. Philosophy of science is a meta-discipline, studying science. It is not the job of philosophers of science to tell scientists how to do science. Stand anywhere on the aim-oriented empiricist side of the watershed, and all this changes. Philosophy of science needs persistently to develop and critically assess new possible metaphysical assumptions, inherent in new possible aims for physics, in an attempt to *improve* the assumption that is accepted, the aim that is pursued. In doing this, philosophy of physics thereby seeks to improve associated *methods* of physics (methods employed in deciding what theories are to be accepted and rejected, along with empirical methods). As we have seen, this activity of attempting to *improve* problematic aims and methods of physics, partly in the light of improving scientific knowledge and understanding, is almost bound to have major implications for physics itself. Indeed, philosophy of physics construed in this way—the enterprise of attempting to improve problematic aims and methods of physics—is a vital, integral part of physics itself, influencing and being influenced by what goes on in the rest of physics, both experimental and theoretical.

These dramatic differences in the whole character of the philosophy of physics and physics itself, and the relationship between the two, all devolve from this key issue: Are there, or are there not, problematic assumptions inherent in the basic aim of physics? Does the basic aim need to evolve, to be improved, or is it unproblematic and fixed?

---

[4] For a more detailed exposition and defence of these points, see Maxwell 2017a and 2017b.
[5] There is here room for slight disagreement. Scientific realists hold that science can acquire knowledge of unobservable entities, whereas instrumentalists hold that scientific knowledge is restricted to observable phenomena.
[6] This holds for such versions of standard empiricism as inductivism, positivism, conventionalism, hypothetico-deductivism, Bayesianism, constructive empiricism, the views of Popper, Kuhn, Lakatos, and others.

## 4. Basic Argument

   I can now state, in a little more detail, the first step of the central argument of this paper. Standard empiricism as characterized above is taken for granted, in one or other version, by most scientists and philosophers of science.[7] It is, however, untenable. Physics in particular and natural science more generally cannot proceed in accordance with its edicts. Given any accepted physical theory, T, endlessly many rivals can be concocted to fit the facts even better than T. Thus, granted that T is Newtonian theory, one rival theory might be Newtonian theory modified so that (a) it has an additional, independently testable and corroborated postulate that successfully predicts phenomena that T cannot predict, and (b) predicts that gravitation will become a repulsive force in the first moment of 2050, so that $F = -Gm_1m_2/d^2$. Another rival might (a) have the additional empirically successful postulate and (b) predict that for gold spheres of mass greater than a thousand tons in outer space, the law of gravitation has the form $F = Gm_1m_2/d^3$. Endlessly many empirically more successful rivals to Newtonian theory can be concocted along these lines. In practice, these are all ignored because they are hopelessly ad hoc, complex, disunified, or non-explanatory. But in only ever accepting *unified* theories, and ignoring endlessly many empirically more successful disunified rivals (of the kind just considered), physics thereby makes a substantial, implicit assumption about the nature of the universe: it is such that all seriously disunified theories are false. The universe is such that empirically more successful disunified rivals to T will make *false* predictions where they clash with the predictions of T. Not only *does* physics make a substantial metaphysical assumption about the universe in persistently accepting only unified theories and ignoring endlessly many empirically more successful disunified rivals; it *must* make some such assumption. If no assumption is made, and theories are assessed impartially with respect to evidence, physics would be overwhelmed by an infinite swamp of empirically successful but horribly disunified theories. Thus, standard empiricism cannot be implemented in scientific practice. If some kind of requirement of simplicity, unity, or explanatoriness is invoked in addition to empirical considerations when it comes to deciding what theories are to be accepted, that means some substantial metaphysical assumption is made about the universe—and that contradicts standard empiricism. Standard empiricism only becomes scientifically viable if it is contradicted in practice![8]

   How is it possible for physics to have been so astonishingly successful if most physicists take for granted the untenable doctrine of standard empiricism? The answer is that they have not taken it too seriously in scientific practice. In practice, considerations that have to do with the simplicity, unity, explanatory, or non-ad-hoc character of a theory are taken very seriously indeed in deciding what is to be accepted and rejected, in addition to empirical considerations. Theoretical physicists seek to develop conservation, invariance, and symmetry principles capable of determining what theories are to be accepted, along with empirical considerations.[9] All these non-empirical requirements a theory must satisfy to be

---

[7] See note 2 and Maxwell 2017a, 73–74.

[8] For more detailed expositions of this refutation of standard empiricism, see Maxwell 1998, chap. 2; 2017a, 69–83; 2017b.

[9] Lorentz invariance, gauge invariance, local gauge invariance, charge, parity and time reversal symmetry, and supersymmetry are all symmetry principles that have played an important role in recent times in theoretical physics. These symmetry principles are, in effect, methodological rules: they specify what features a theory must have if it is to be a candidate for acceptance. At the same time, they are methodological principles that have a fallible, conjectural aspect to them: they may need to be rejected (and some have been rejected). This

accepted in effect appeal to implicit metaphysical principles.[10] Standard empiricism is quietly and radically violated in scientific practice, without this being acknowledged. It is this that has made possible progress in physics—and progress in science more generally.[11] As Einstein once declared, "If you want to find out anything from the theoretical physicist about the methods they use, I advise you to stick closely to one principle: don't listen to their words, fix your attention on their deeds" (1973, 270).

Philosophers of science, however, have taken standard empiricism much more seriously. The whole way the discipline is conceived and pursued, the way its relationship to science is conceived and maintained, presupposes standard empiricism. All the diverse, well-known doctrines about science that philosophers of science have developed are versions of standard empiricism, as I have already indicated. It is this that accounts for the scientific poverty of academic philosophy of science, its irrelevance to science, so cruelly pointed out by scientists themselves. The intellectual activity of developing, exploring, and critically assessing possible aims for physics, possible metaphysical assumptions inherent in these aims—an activity that constitutes such a vital, integral part of physics if one stands on the aim-oriented empiricist side of the watershed—does not make sense, and is not a viable or coherent endeavour for philosophy of physics if one stands on the standard empiricist side of the watershed. On this side of the watershed, the basic aim of physics is fixed; it does not make sense to try to improve it. Physics does not make problematic metaphysical assumptions concerning the physical comprehensibility of the universe, at least not in the context of justification, and so there is no philosophical task of trying to improve these assumptions. Such assumptions may be made in the context of discovery, but that, viewed from this side of the watershed, is *physics*, not *philosophy of physics*. Standard empiricism thus cripples the philosophy of science and renders it more or less irrelevant from the standpoint of science itself—as scientists themselves have attested.

Put in Kuhnian terms, philosophy of science has long been in a state of crisis, at least since the heady days of logical positivism. A revolution is long overdue. This would involve a radical change of paradigm, from standard to aim-oriented empiricism. The outcome would be a transformation in the nature of the discipline. Both science and philosophy of science would change, and the relationship between the two would change as well. As I have put it elsewhere, science and philosophy would be brought together to create a modern version of natural philosophy.[12]

---

accords with aim-oriented empiricism but is starkly at odds with standard empiricism (which does not really have a role for *evolving* methodological rules).

[10] In demanding that a theory satisfy such a requirement to be accepted, physics implicitly accepts the thesis that the universe is (or the phenomena are) such that a theory, in order to be near enough to the truth to be acceptable, must satisfy the requirement. In other words, the universe is (or the phenomena are) such that all theories that fail to satisfy the requirement are false.

[11] As a result of taking standard empiricism as seriously as they do, however, physicists cannot take up, in the explicit, sustained way that is required, the articulation and critical assessment of metaphysical conjectures implicit in the symmetry principles they propose, consider, and adopt. For to do so would be to violate the precepts of standard empiricism all too blatantly. For an informal and entertaining account of the confusion that results, see Hossenfelder 2018. See also Maxwell 2017a, chap. 5.

[12] See Maxwell 2017a. See also Maxwell 1984, chap. 9; 1998; 2004.

## 5. Recent Work on the Metaphysics of Science

In recent years the metaphysics of science has become a hot topic in the philosophy of science. Does this work acknowledge that metaphysics plays a role in science only in the context of discovery, a metaphysical thesis being acceptable only if it is compatible with current scientific knowledge, there being no hint of "the basic argument" of section 4 for aim-oriented empiricism, some version of standard empiricism thus being presupposed? Or is some version of "the basic argument" propounded, aim-oriented empiricism thus being defended, there being a full recognition of the point that the aims and methods, and thus the metaphysics too, of science evolve with evolving scientific knowledge? In other words, on which side of the watershed between standard and aim-oriented empiricism does this body of work stand? Let us see.

Dilworth 2007 is, in some respects, a pioneer as far as this relatively recent body of work is concerned—the book was first published in 1996. There is a clear recognition that metaphysical principles play an important role in science. However, despite several references to Maxwell 1984, where aim-oriented empiricism is expounded and defended, there is no hint in Dilworth's book of "the basic argument" for, and key tenets of, aim-oriented empiricism. For a critical appraisal see Maxwell 2009.

Ladyman et al. 2007 decisively criticises that enterprise of analytic philosophy that seeks to do metaphysics independently, or in ignorance, of modern physics. It is clearly recognized that the task of scientific metaphysics is to provide a basis for the unification of two or more accepted fundamental theories of physics, all such proposals being conjectural and likely to be false. The account of unification is, however, unsatisfactory—it takes Kitcher 1976 and 1981 for granted.[13] There is no hint of the conception of unification that is required: see Maxwell 1998, chaps. 3–4, and 2004, 160–74. Much more seriously, there is no hint of "the basic argument" of section 4 above for aim-oriented empiricism. Far from explicating something like the hierarchical meta-methodology of aim-oriented empiricism (see below), designed to subject the metaphysics of physics to sustained critical scrutiny and attempted improvement, Ladyman et al. 2007 actually states at one point that "there is no such thing as 'scientific method'" (27). Like Dilworth, Ladyman and colleagues do not take up "the new task for philosophy of science" and do not argue that it should be done.

Chakravartty (2007) is concerned to articulate a version of scientific realism that does justice both to modern science and to the critics of realism, but all within the framework of standard empiricism, which is taken for granted throughout. There is not a hint of "the basic argument," nor of the need for a new hierarchical meta-methodology for physics designed to facilitate the improvement of the metaphysics of physics as science proceeds.

Lange (2009) is concerned to clarify what it is in existence that renders a law of nature, even though contingent, nevertheless in some sense "necessary," and distinct from a true accidental generalization.[14] Standard empiricism is presupposed without discussion.

Morganti (2013) seeks to develop a metaphysics for science that steers a middle course between scientific naturalism and philosophical a priorism. There is no hint of aim-oriented empiricism or the new task for philosophy of science.

In their excellent introduction, Mumford and Tugby begin by stating that "science can only exist in an ordered, patterned world, and it is argued that the core aim of the metaphysics of science is to investigate the nature of that order" (2013, 3). This promising beginning does not, however, lead on to the point that this metaphysical thesis of patterned order is a problematic presupposition of physics, very likely to be false in the specific form adopted at any given stage in the development of physics, a new hierarchical methodology for physics being required to subject the thesis to sustained criticism and attempted improvement.

---

[13] For a decisive criticism of Kitcher 1976 and 1981, see Maxwell 1998, 62–68.
[14] For another view on this issue, see Maxwell 1968. See also Maxwell 2019a, chap. 1.

Kincaid, Ladyman, and Ross 2013 is a collection of essays devoted to the idea that metaphysics must be based on modern science. The book contains contributions from Harold Kincaid, Anjan Chakravartty, Paul Humphreys, Andrew Melnyk, Daniel Dennett, James Ladyman and Don Ross, Mark Wilson, Michael Friedman, and Jenann Ismael. As Kincaid mentions in the introduction, many of these contributors have recently published books on scientific metaphysics. It is thus highly significant that nowhere do we find, in this collection, any mention of "the basic argument" that leads one to acknowledge that physics makes a highly problematic metaphysical presupposition concerning underlying unity in nature, a new meta-methodology being required to subject this presupposition to sustained scrutiny and attempted improvement. The first steps towards "the new task for philosophy of science" are nowhere taken.[15]

What all these authors ignore is a central source for the metaphysics of physics that comes, not from physical theory, but from the *methods* of physics—specifically that methodological rule that asserts: in order to be acceptable, a fundamental physical theory must be (sufficiently) *unified*.[16] It is the persistent acceptance of unified theories only, when endlessly many empirically more successful disunified rivals are available, that commits physics to the metaphysical presupposition that, at the very least, the universe is such that all disunified theories are false. Recognition of this point constitutes the key step towards adopting the hierarchical methodology of aim-oriented empiricism (see below), and the new task for philosophy of science. I have developed this argument in detail in a series of publications since 1974--see Maxwell (1974; 1984; 1998; 2004; 2005) and elsewhere[17]—and yet none of the above publications refers to this body of work, apart from Dilworth 2007.[18] On the other hand, those who have read these publications of mine speak well of them: see, for example, Kneller (1978, 80-87 and 90-91); Longuet-Higgins (1984); Smart (2000); Cory (2000); McHenry (2000); Roush (2001); Muller (2004); and MacIntyre (2009).

**6. An Improved Version of Aim-Oriented Empiricism**

Aim-oriented empiricism has one crucial advantage over standard empiricism. Whereas the latter denies, falsely, that physics makes a substantial metaphysical assumption about the nature of the universe, aim-oriented empiricism, correctly, acknowledges that such an assumption is made, recognizes that it is a mere conjecture all too likely to be false in the specific version accepted at any stage in the development of physics, and throws it open to sustained critical scrutiny in an attempt to improve it. Aim-oriented empiricism, as formulated briefly above, still faces problems, however. What possible justification can there be for physics to presuppose that the universe is physically comprehensible in the sense that the true physical theory of everything is *unified*? What does it mean, in any case, to say of a theory that it is unified? How can metaphysical theses about the comprehensibility of the universe be *improved*? What if the universe is not physically comprehensible? Does that just mean that physics becomes impossible?

---

[15] For a much more detailed critical appraisal of the above-mentioned works of all these authors, apart from Lange 2009, along the lines indicated here, see Maxwell 2019a, chap. 4.

[16] Mumford and Tugby (2013, introduction) come closest to recognizing this crucial point.

[17] See especially Maxwell (2017b; also 2017a and 2017c).

[18] Additional recent works on the metaphysics of science I have examined that take standard empiricism for granted and fail even to mention, let alone discuss, aim-oriented empiricism dating back to Maxwell 1974 include: Ellis 2001; Lowe 2006; Bird 2007; M. O'Rourke, M. H. Slater, A. Borhini, P. Godfrey-Smith, N. Latham, R. Sorensen, A. C. Varzi, M. Devitt, B. Nany, N. E. Williams, B. Glymour, N. G. Rheins, J. K. Crane, R. Sandler, and K. Vihvelin in Campbell et al. 2011; Trout 2016; S. Yudell, K. Brading, M. Strevens, C. K. Waters, K. Stanford, J. Saatsi, and M. Thomas-Jones in Slater and Yudell 2017.

In order to solve these problems, the version of aim-oriented empiricism briefly formulated above needs to be radically developed and improved. Elsewhere, I have developed and improved the view in some detail along these lines, and I have shown in detail how it solves these (and related) problems, so here I will be brief.

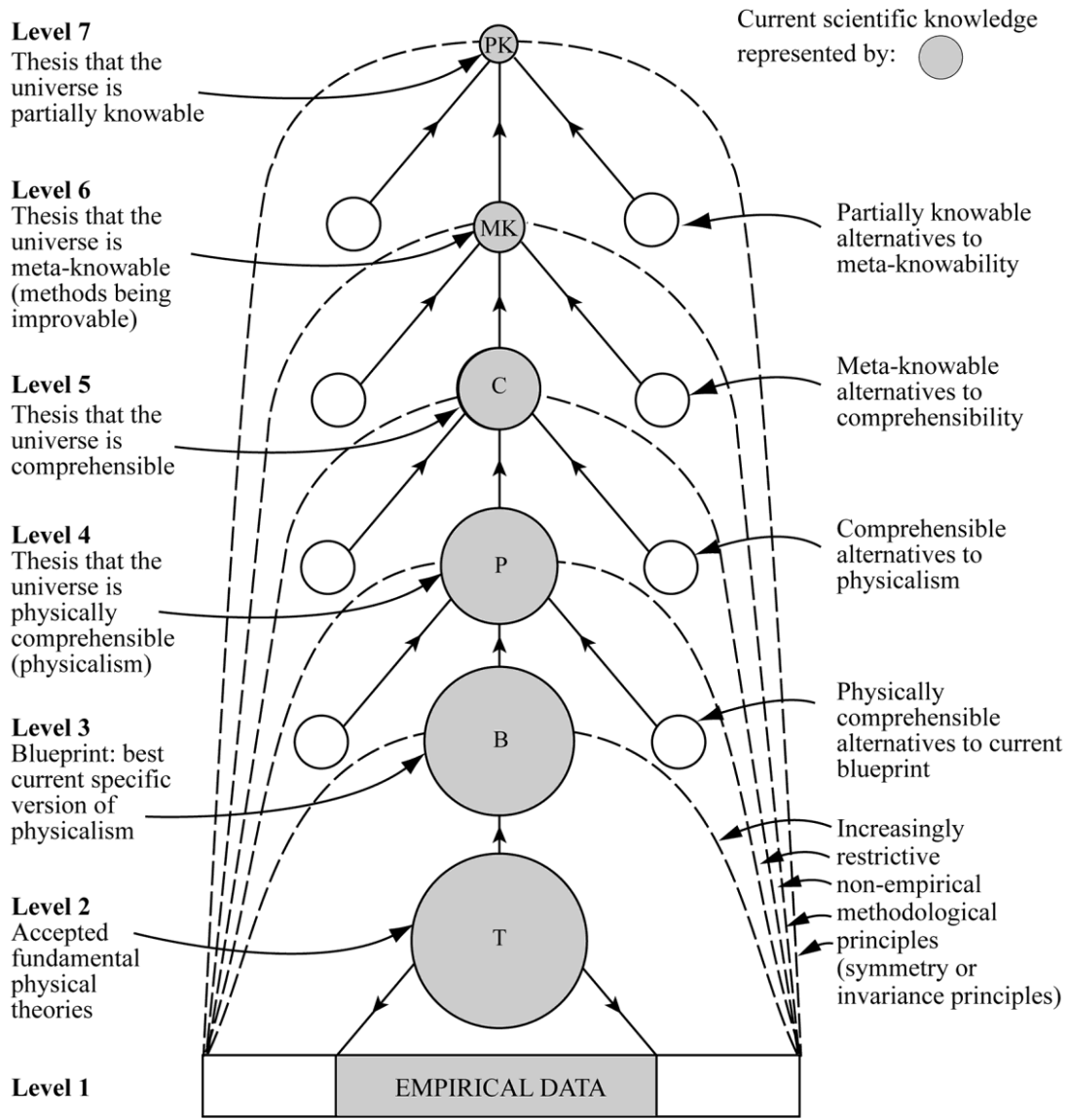

Figure 1. Aim-oriented empiricism

Aim-oriented empiricism needs to be formulated in such a way that physics makes, not just one metaphysical assumption, but a hierarchy of such assumptions: see figure 1. As one goes up the hierarchy, assumptions become increasingly insubstantial, and so increasingly likely to be true, and increasingly such that their truth is required for science, or the pursuit of knowledge, to be possible at all.

At the top there is the thesis that the universe is such that we can continue to acquire knowledge of our immediate environment sufficient to make life possible. Next down, at level 6, there is the thesis that the universe is such that we can formulate a conjecture about it that enables us to improve our methods for improving knowledge. These two assumptions are accepted permanently, not because we have any reason to hold them to be true, but because, granted we seek to improve knowledge of truth, it can only help and cannot hinder the search for knowledge. Next down, at level 5, there is the thesis that the universe is comprehensible in some way or other. There is some one kind of explanation for all phenomena. Everything that occurs does so because a society of gods wills it, or one God wills it, or because it goes towards the realization of some cosmic goal, or because it is in accordance with some cosmic computer programme, or because it occurs in accordance with some unified pattern of physical law—or because some other entity is, in some way, responsible for everything that occurs. This has the merit of exemplifying the thesis one up in the hierarchy: if the thesis is true, then we may progressively improve our methods for the improvement of knowledge (at levels 1 and 2), by honing in on that version of the thesis that best simulates the growth of empirical knowledge. Next down, at level 4, there is the thesis that everything occurs in accordance with a unified pattern of physical law. This version of the thesis that the universe is comprehensible—*physicalism*, as it may be called—has been astonishingly fruitful in stimulating the growth of knowledge in physics, at levels 1 and 2. All the great developments in theoretical physics since Kepler and Newton exemplify theoretical *unification*, and thus this level-4 thesis. Newtonian theory unifies Kepler and Galileo, terrestrial and astronomical phenomena. Maxwellian electrodynamics unifies the electric and magnetic fields, and unifies light, infra-red, radio, and ultra-violet rays, X-rays, and gamma rays in revealing they are all electromagnetic waves of different wavelengths. Special relativity brings greater unity to classical electrodynamics, partially unifies energy and mass via the equation $E = mc^2$, and partially unifies space and time by means of Minkowski space-time. General relativity unifies space-time and gravitation. Quantum theory and the atomic theory of matter unify a vast array of laws concerning the physical and chemical properties of matter, and the way matter interacts with light. Quantum electrodynamics unifies quantum theory, classical electrodynamics, and special relativity; quantum electroweak theory unifies quantum electrodynamics and the weak force; quantum chromodynamics brings unity to the physics of hadrons. All these theoretical developments enormously enhance the predictive scope of theory *and at the same time* bring ever greater unity to physical theory, thus drawing ever closer to capturing the level-4 metaphysical thesis in the form of a testable, unified physical theory of everything. In other words, this level-4 thesis has been astonishingly empirically fruitful in supporting an immensely empirically progressive research programme—the whole enterprise of theoretical physics since Newton. At level 3 there is the best available specific version of physicalism; today, this may be held to be string theory. At level 2, there are the accepted fundamental theories of physics, general relativity and the standard model—the quantum field theory of fundamental particles and the forces between them. And at level 1 there is empirical phenomena, the low-level empirical laws of experimental results.

 This hierarchical version of aim-oriented empiricism provides a framework of relatively unproblematic assumptions and associated methods—aims and methods—at levels 7 and 6, accepted and adopted permanently, within which increasingly problematic aims and methods, as we go from level 5 to level 3, may be critically assessed, developed, and, we may hope, improved. At levels 5 to 3, that thesis is accepted which (a) best accords with the accepted thesis above it in the hierarchy and (b) supports, or promises to support, the most empirically progressive research programme at levels 1 and 2. Theses at levels 7 and 6 are, we may conjecture, true; as we descend from level 6 to level 3, we move from truth to falsity. The hierarchy concentrates criticism and attempts at improvement where they are likely to be most fruitful, low down in the hierarchy, and at the same time provides a fixed framework, at levels 7 and 6, that restricts ways in which theses, lower down in the hierarchy, can be modified to those ways likely to be most fruitful from the standpoint of progress in physics.

This fixed framework facilitates something like positive feedback between improving scientific knowledge and improving knowledge about how to improve knowledge—improving assumptions and associated methods, in other words. As we improve our knowledge of nature, we correspondingly improve the nature of science. Everyone would accept that this goes on at the experimental level: new knowledge leads to new experimental techniques, new instruments such as the microscope, the telescope, the particle accelerator. Aim-oriented empiricism specifies how this can go on at the theoretical level, new theories leading to new metaphysical ideas, and new metaphysical ideas leading to the discovery of new theories.[19] This has occurred in the history of physics—or we would still be stuck with Aristotle—but it would proceed in a much more explicit fashion, in a way likely to be much more fruitful, if aim-oriented empiricism were explicitly to be put into scientific practice.[20]

This hierarchical version of aim-oriented empiricism depicts the fusion of physics and the philosophy of physics. It is vital, for the intellectual rigour and success of the enterprise, that each should influence, and be an integral part of, the other.[21]

I might add that the level-3 thesis and associated methods (level-3 aims and methods) are methods for the discovery and acceptance of level-2 theories;[22] the level-4 aim and methods are meta-methods for the discovery and acceptance of level-3 theses and associated methods; and so on as we go up the hierarchy. Methods associated with the level-7 thesis are meta-meta-meta-meta-methods! It is this meta-structure of the methods of this hierarchical view that makes it possible for it to facilitate positive feedback between improving knowledge and improving metaphysical theses and associated methods—improving knowledge about how to improve knowledge, in other words.

Elsewhere, I have shown that aim-oriented empiricism is both required, and sufficient, to solve the problem of induction (see Maxwell 2017b, esp. chap. 9). This is a decisive argument in support of the view, and against all versions of standard empiricism. (Centuries of effort, ever since David Hume, to solve the problem presupposing versions of standard empiricism, have not met with success.)

## 7. Broader Implications

So far I have indicated a new task for the philosophy of physics: convince physics that aim-oriented empiricism needs to be put into practice, and then collaborate with theoretical physicists in doing just that. The scope of the above argument can, however, be enlarged so that it becomes applicable, not just to physics, but to the whole of natural science.

In order for considerations analogous to the ones spelled out above in connection with physics to apply to some other science, that science must have permanently problematic *aims*, so that its aims need sustained critical scrutiny in an attempt to improve them; furthermore, aims are improved in such a way that associated *methods* need to be improved as well. If these conditions obtain, then the activity of improving aims and methods—the philosophy of science, in other words—needs to be pursued as an integral, influential part of the science itself. Wherever aims and methods need to evolve with evolving science, a version of the hierarchical structure of aim-oriented empiricism needs to be put into scientific practice. These conditions hold in what follows. I have five points to make.

---

[19] For a detailed account of the fallible but rational method of discovery that aim-oriented empiricism provides, and other fruitful implications for theoretical physics that the view has, see Maxwell 2017a, chap. 5. See also Maxwell 1993.
[20] This is spelt out in detail in Maxwell 2017a, chap. 5.
[21] See Maxwell 2017a and 2017b. See also Maxwell 1984, chap. 9; 1998; 2004.
[22] Elsewhere I have demonstrated that this hierarchical view provides physics with a rational, if fallible, method for the discovery of new physical theories: see Maxwell 1993; 1998, 219–23; 2004, 191–205; and especially 2017a, chap. 5.

The first point to note is that physics is the fundamental natural science, aspects of which are presupposed directly or indirectly by all other branches of natural science. Thus chemistry, cosmology, astronomy, and geology all presuppose aspects of physics; biology and neuroscience presuppose aspects of chemistry (and probably, in a direct but looser way, aspects of physics as well). The above argument, ostensibly about physics only, is actually about the whole of natural science.

Secondly, the above argument applies in a much more direct way to all other branches of natural science. It is not just physics that makes problematic assumptions inherent in its basic aim; this is true of all the branches of natural science. Chemistry, molecular biology, genetics, neuroscience, geology, astronomy, climate science: all make assumptions that are composed partly (a) of items of knowledge from some more fundamental science and partly (b) of conjectures about what there is for the science in question to seek to discover and transform into scientific knowledge. The latter component (b) is inherently and particularly problematic, in that it concerns the domain of our ignorance, that which we conjecture, do not yet know, but may come to transform into scientific knowledge. Genetics seeks to discover what genes do and how they do it; cell biology seeks to improve knowledge and understanding of what goes on in the cell; neuroscience seeks to improve knowledge and understanding of how what goes on in the brain is related to what happens to the person, what the person does, thinks, perceives, remembers. And so on. No branch of science is entirely ignorant of that which remains to be discovered; there are always more or less inadequate conjectures or intimations about what exists to be discovered, and what can be discovered by current methods. Thus, for each branch of natural science, considerations that arise in connection with theoretical physics arise here too: each branch needs to make explicit problematic *aims* of the discipline, problematic *assumptions* inherent in these aims, so that they may be critically assessed, developed, and, we may hope, improved as the science proceeds. More or less specific, problematic aims, and associated methods, of each scientific discipline need to be articulated, critically assessed, developed, and improved as an integral part of the discipline itself. Each discipline, in other words, needs to implement its own version of aim-oriented empiricism—the philosophy of the discipline forming an important, integral, influential part of the discipline itself. It may well be an unnecessary extravagance for a science such as geology or astronomy to put a version of aim-oriented empiricism into practice that has a hierarchical structure of *seven* levels of assumptions—like the seven levels of the version of aim-oriented empiricism indicated above, applied to theoretical physics. On the other hand, the *two* levels of standard empiricism, evidence and theory, are definitely insufficient. Each scientific discipline needs to acknowledge and represent at least *three* levels of sustained discussion: evidence, theory, and *aims*—the latter including problematic assumptions inherent in aims. Once a science is constituted and pursued in this fashion, it brings together, and *fuses*, the science and its philosophy of science.

Thirdly, there is another, very important way in which the above considerations apply to all the diverse branches of natural science, and to natural science as a whole. The aims of science and the sciences are inherently problematic, not just because they have problematic *metaphysical* and other factual assumptions inherent in them; they are problematic because they have *value* assumptions in them, and assumptions concerning *social use*. The basic intellectual aim of science is not just *truth*, as standard empiricism assumes; nor is it just *explanatory truth* (truth presupposed to be inherently explanatory) as the version of aim-oriented empiricism expounded above in sections 2 and 5 presupposes: more generally, it is to improve knowledge of *valuable truth*, truth deemed to be significant, useful, important, of interest or of value in some way, explanatory truth being a special case of valuable truth. And valuable truth is sought so that, once acquired, it can be used by people, either to enhance their knowledge and understanding of the world around them for its own sake or to attain other human goals by means of technological applications and in other ways. There are, in other words, problematic humanitarian, social, even political assumptions concerning the human use of science inherent in the aims of science.

In order to become an item of scientific knowledge, it is not sufficient that a result is new and very well established: it must reach a certain threshold of significance, of value. One cannot contribute to science by counting gravel on paths or leaves on trees—not unless this is a part of a broader, more significant research programme. A science that accumulated a vast store of well-established but irredeemably trivial facts would not thereby be said to be making splendid progress. It is inevitable that values play a role in deciding what does and what does not become a part of scientific knowledge. The domain of fact is infinite; some aspects of the world around us must be selected out as the significant aspects for science to concentrate on. And it is desirable that science should do this: we want science to improve our knowledge about what matters or is of genuine interest or use; we do not want knowledge of irredeemable trivia.

It is of course vital to appreciate that value judgements influence what enters the domain of scientific knowledge in a way that is entirely different from the way metaphysical theses concerning the comprehensibility and knowability of the universe influence acceptance of physical theory, as depicted above in sections 2 to 5. There, metaphysical assumptions influence judgements about *truth*. Values, however, ought not to influence judgements about *truth*, since we have no reason to hold that just because it would be good for something to be true, it is likely to be true, or just because it would be bad, it is likely to be false.[23] Values play a role in determining whether or not a given result should enter scientific knowledge—whether it should be published in a scientific journal, for example in influencing decisions about whether the result is sufficiently *significant* or *important* to be published, to become a part of scientific knowledge.

Assumptions concerning value and social use are inherent in the aims of science, and in the aims of diverse sciences, in contexts of both discovery and justification. Here, the context of discovery is, perhaps, the more important, for it is in this context that decisions are made about the priorities of research, what research projects to fund and not to fund.

Value assumptions inherent in the aims of science, and in the aims of specific sciences, are intrinsically and permanently problematic.[24] Of value to whom? In what kind of way? Of value in the immediate future, or in fifty or a hundred years' time? Who is to decide, and how? In addition to the obvious, general problems that arise in deciding questions of value, there are some very difficult questions that are specific to science. First, in attempting to choose the best aims for scientific and technological research, we seek those aims that are (a) possible to discover and (b) desirable or of value to discover. Both (a) and (b) are inherently problematic, (b) especially. It is notoriously difficult to anticipate future uses and misuses of scientific discoveries, ways in which discoveries will turn out to be used for benefit and for harm.

---

[23] If values do influence judgements concerning epistemological merit, then this influence should be *negative*: the more important it is for a result to be true, the more we desire it to be true, the more there is at stake, then the harsher and more severe should be the experimental examination of the result, the higher our standards for acceptance. This is especially the case if lives are at stake, as in the use of new drugs.

[24] This is a basic theme of Maxwell 1984: see in particular chap. 5. It is also a theme of Kitcher 2001. Kitcher informed me that in writing his book he was influenced by his reading of Maxwell 1984.

Second, there is the permanent problem that science can be of value in two very different ways: of value intellectually or culturally, in enhancing our knowledge and understanding of aspects of the world around us, and of value technologically or practically, in enabling us to achieve other things of value: food, health, shelter, transport, communications, and all the benefits of the modern world. How are these two very different kinds of value to be weighed up against each other? How is the value of pure science to be weighed against the value of saving lives and alleviating human suffering? Scientists sometimes argue that pure science is justified because in the end it results in beneficial technology, but this is by no means always the case. Research into cosmology or the unification of general relativity and the standard model is unlikely, however successful, to have practical consequences.

Third, there is the problem that modern science is very expensive, but those who stand most in need of the products of science are the poor of the earth. Ideally, science ought to serve the best interests of humanity. That would require that the priorities of research reflect the priorities of human need, the needs above all of the poor of the planet. It is all but inevitable, however, that priorities of research will reflect the interests of scientists themselves, and the interests of those who pay for science: wealthy governments, wealthy corporations, wealthy populations, and the military of wealthy nations. The tendency of research to reflect the interests of those who pay for it must inevitably pull priorities of research away from responding to the most urgent and important interests of humanity. Does not the world spend far too much on military research, and not nearly enough on technology needed to decrease $CO_2$ emissions rapidly, and thus put a stop to global warming? Do the priorities of medical research reflect the health needs of the poor in developing countries or the health needs of those living in wealthy countries, as one might expect? These kind of considerations ensure that value assumptions, and assumptions about social use, inherent in the aims of science, will be permanently and profoundly problematic.

Because the aims of science, in the contexts of both discovery and justification, have profoundly problematic assumptions inherent in them concerning values and social use, it is, as before, vital that actual and possible aims be imaginatively articulated and critically assessed, in a sustained way, as an integral part of science, in an attempt to improve them. And that means that the philosophy of science needs to be an essential, influential part of science here too.

This context of values and social use is, however, in one crucial respect, different from contexts discussed previously. Scientists and philosophers of science need to participate in rational—that is, imaginative and critical—exploration of actual and possible aims of science, in an attempt to improve them. They ought not, however, to be the only participants in this discussion. Scientists and philosophers of science cannot be in a position to decide for the rest of humanity what is of value and what are the world's most important needs. Non-scientists must be able to contribute to the discussion of aims for science on an equal footing with scientists and philosophers of science, especially when it comes to decisions about value and social use—decisions about priorities of research that stem from values and social use.

The intellectual/institutional structure of science needs to be modified to accommodate both scientists and non-scientists in the collaborative discussion of aims and priorities of scientific research. We need new journals, new public committees, new public lectures and debates. Science journalists—and philosophers of science—need to help create and keep alive this public discussion.

Governments, funding bodies, commercial interests, charities, and scientists themselves will of course continue to make decisions about what research to support, or do. The all-important point is that this decision-making should be bathed in the results of good public exploration of actual and possible research aims, so that decision-making may be both improved and assessed by its means.

Throughout science, in short, in all contexts, the intellectual domain of science needs to consist of, not just *two* domains, evidence and theory, but as a bare minimum, *three*: evidence, theory, and aims. Contributions to aims may be as important as contributions to theory or evidence. It should be possible to win a Nobel Prize as a result of a really important and original contribution to the domain of *aims*.

## 8. Conclusion

We philosophers of science have a major task on our hands. We need to convince scientists—and governments and the public—that a revolution in the nature of science is urgently needed. In a wide range of contexts, problematic *aims* of science need sustained imaginative and critical discussion in an attempt so to develop science that it comes to serve the very best interests of humanity. And furthermore, philosophers of science need to participate in this crucial scientific activity of discussing, and attempting to improve, problematic aims of science. Philosophers of science need to become scientists; and some scientists at least need to become philosophers.

Elsewhere, I have argued that there are even broader repercussions of the above argument. I have argued that the hierarchical, aim-improving methodology of aim-oriented empiricism has implications, when generalized, not just for science, but for all worthwhile human endeavours with problematic aims.[25] I have argued that it is not just science and its aims that need to be transformed; this is true of the whole academic enterprise. The basic intellectual aim of academic inquiry as a whole should be, not just to acquire specialized knowledge, but rather to seek and promote social wisdom—wisdom being the capacity, the active endeavour, and the desire to achieve what is of value in life for oneself and others, thus including knowledge, understanding, and technological know-how, but much else besides.[26]

*nicholas.maxwell@ucl.ac.uk*

## References

Bird, Alexander. 2007. *Nature's Metaphysics: Laws and Properties*. Oxford: Clarendon Press.
Campbell, Joseph, Michael O'Rourke, and Matthew Slater, eds. 2011. *Carving Nature at Its Joints: Natural Kinds in Metaphysics and Science*. Cambridge, Mass.: MIT Press.
Chakravartty, Anjan. 2007. *A Metaphysics for Scientific Realism*. Cambridge: Cambridge University Press.
Cory, F. J. 2000. Review of Nicholas Maxwell, *The Comprehensibility of the Universe*. *International Philosophical Quarterly* XL, no. 4:517–18.
Dilworth, Craig. 2007. *The Metaphysics of Science*. Dordrecht: Springer.
Einstein, Albert. 1973. *Ideas and Opinions*. London: Souvenir Press.

---

[25] See Maxwell 1984, chaps. 5–11; 2000; 2004, chaps. 3–4; 2014; 2017a, chap. 8; 2017c, prologue and chap. 10; 2019b.

[26] See the works referred to in the previous note.

Ellis, Brian. 2001. *Scientific Essentialism*. Cambridge: Cambridge University Press.
Hossenfelder, Sabine. 2018. *Lost in Math: How Beauty Leads Physics Astray*. New York: Basic Books.
Kincaid, Harold, James Ladyman, and Don Ross, eds. 2013. *Scientific Metaphysics*. Oxford: Oxford University Press.
Kitcher, Philip. 1976. "Explanation, Conjunction and Unification." *Journal of Philosophy* 73:207–12
———. 1981. "Explanatory Unification." *Philosophy of Science* 48:507–31.
———. 2001. *Science, Truth and Democracy*. Oxford: Oxford University Press.
Kneller, George. 1978. *Science as a Human Endeavor*. New York: Columbia University Press.
Ladyman, James, and Don Ross, with David Spurrett and John Collier. 2007. *Every Thing Must Go*. Oxford: Oxford University Press.
Lange, Marc. 2009. *Laws and Lawmakers: Science, Metaphysics and the Laws of Nature*. Oxford: Oxford University Press.
Longuet-Higgins, Christopher. 1984. "For Goodness Sake." *Nature* 312:204.
Lowe, E. Jonathan. 2006. *The Four-Category Ontology: A Metaphysical Foundation for Natural Science*. Oxford: Oxford University Press.
MacIntyre, Aladair. 2009. "The Very Idea of a University." *British Journal of Educational Studies* 57, 4:358.
Maxwell, Nicholas. 1968. "Can There Be Necessary Connections Between Successive Events?" *British Journal for the Philosophy of Science* 19, no. 1:1–25.
———. 1974. "The Rationality of Scientific Discovery: Part I." *Philosophy of Science* 41:123–53.
———. 1984. *From Knowledge to Wisdom: A Revolution in the Aims and Methods of Science*. Oxford: Blackwell.
———. 1993. "Induction and Scientific Realism: Part Three: Einstein, Aim-Oriented Empiricism and the Discovery of Special and General Relativity." *British Journal for the Philosophy of Science* 44, no. 2:275–305.
———. 1998. *The Comprehensibility of the Universe: A New Conception of Science*. Oxford: Oxford University Press.
———. 2000. "Can Humanity Learn to Become Civilized?" *Journal of Applied Philosophy* 17:2-44.
———. 2004. *Is Science Neurotic?* London: Imperial College Press.
———. 2005. "Popper, Kuhn, Lakatos, and Aim-Oriented Empiricism." *Philosophia* 32:181–239.
———. 2007. *From Knowledge to Wisdom: A Revolution for Science and the Humanities*. 2nd extended edition. London: Pentire Press.
———. 2009. Review of Craig Dilworth, *The Metaphysics of Science. International Studies in the Philosophy of Science* 23, no. 2:228–32.
———. 2013. "Has Science Established That the Cosmos Is Physically Comprehensible?" In *Recent Advances in Cosmology*, edited by Anderson Travena and Soren Brady, 1–56. New York: Nova Science.
———. 2014. *How Universities Can Help Create a Wiser World*. Exeter: Imprint Academic.
———. 2017a. *In Praise of Natural Philosophy*. Montreal: McGill-Queen's University Press.
———. 2017b. *Understanding Scientific Progress*. St. Paul: Paragon House.
———. 2017c. *Karl Popper, Science and Enlightenment*. London: UCL Press (free online).
———. 2019a. *The Metaphysics of Science and Aim-Oriented Empiricism: A Revolution for Science and Philosophy*. Cham: Springer.

———. 2019b. *Science and Enlightenment: Two Great Problems of Learning*. Cham: Springer.
McHenry, Leemon. 2000. Review of Nicholas Maxwell, *The Comprehensibility of the Universe*. *Mind* 109:162–66.
Morganti, Matteo. 2013. *Combining Science and Metaphysics*. London: Palgrave Macmillan.
Muller, F. A. 2004. "Maxwell's Lonely War." *Studies in History and Philosophy of Modern Physics* 35:109–10 and 117.
Mumford, Stephen, and Matthew Tugby, eds. 2013. *Metaphysics and Science*. Oxford: Oxford University Press.
Popper, Karl. 1959. *The Logic of Scientific Discovery*. London: Hutchinson.
Roush, Sherrilyn. 2001. Review of Nicholas Maxwell, *The Comprehensibility of the Universe*. *Philosophical Review* 110:85–87.
Slater, Matthew, and Zanja Yudell, eds. 2017. *Metaphysics and the Philosophy of Science*. Oxford: Oxford University Press.
Smart, John Jamieson Carswell. 2000. Review of Nicholas Maxwell, *The Comprehensibility of the Universe*. *British Journal for the Philosophy of Science*. 51: 907-11.
Trout, J. D. 2016. *Wondrous Truths*. Oxford: Oxford University Press.